\documentclass{sf2a-conf}
\usepackage{graphicx,natbib}

\begin{document}
\TitreGlobal{SF2A 2008}
\title{The Galactic bulge as seen in optical surveys}
\author{Reyl\'e, C.} \address{Observatoire de Besan\c{c}on, Institut UTINAM, Universit\'e de France-Comt\'e, BP 1615, 25010 Besan\c{c}on Cedex, France}
\author{Marshall, D. J.} \address{D\'epartement de Physique et Centre Observatoire du Mont M\'egantic, Universit\'e Laval, Qu\'ebec, G1K 7P4, Canada}
\author{Schultheis, M.$^1$} 
\author{Robin, A. C.$^1$} 
%
%
\setcounter{page}{237} 


\maketitle
\begin{abstract}
The bulge is a region of the Galaxy of tremendous interest for understanding galaxy formation. However measuring photometry and kinematics in it raises several inherent issues, such as severe crowding and high extinction in the visible. Using the Besan\c{c}on Galaxy model and a 3D extinction map, we estimate the stellar density as a function of longitude, latitude and apparent magnitude and we deduce the  possibility of  reaching and measuring bulge stars with Gaia.
We also present an ongoing analysis of the bulge using the Canada-France-Hawaii Telescope.
\end{abstract}
%
\section{Introduction}

Observing towards the bulge gives measures for a large number of individual stars that are the tracers of the Galactic formation and evolution. A detailed study of the bulge is necessary to understand its structure, dynamics and formation. The bulge of the Milky Way is the only one where the parameters in the six dimensions of phase space can be determined, as well as elemental abundances, on a star by star basis. Its detailed observation is thus very useful to test different scenarii of galactic bulges formation.

The bulge has been explored by a few ground based surveys, for example DENIS, 2MASS and microlensing surveys (OGLE, MACHO, DUO, EROS-2 and MOA) in the visible and the infrared, and from space by Spitzer in the mid infrared. The interstellar matter distributed in the plane is a major obstacle to the observation of the stars in the direction of the bulge and the Galactic center, particularly in the visible.

Gaia will give unprecedented view of the Galaxy, but it is often stated that, due to the wavelength coverage of instruments, it will be hardly usable for bulge studies. However, estimations of what Gaia will effectively see in the bulge is worthwhile to do before launch. We have investigated this question using a population synthesis model and a realistic 3D map of the extinction in the inner Galaxy. Studies of the bulge from ground before Gaia are also possible. We present an on-going survey of the bulge undertaken with Megacam aimed at producing good photometry and proper motions for a substancial number of clump stars.

\section{Interstellar extinction}
\label{section1}

Extinction is so clumpy in the Galactic plane that it determines for a great part the number density of stars, more than any other large scale galactic structure. Thus, it is possible to extract information about the distribution of the extinction from photometry and star counts. \cite{Marshall2006} have shown that the 3D extinction distribution can be inferred from the stellar colour distributions in the 2MASS survey. Using stellar colours in $J-K_S$ as extinction indicators and assuming that most of the Galaxy model prediction deviations on small scales from observed colours arises from the variation of extinction along the line of sight, they built a 3D extinction map of the galactic plane. The resulting 3D extinction map furnishes an accurate description of the large scale structure of the disc of dust. 

\section{The bulge density law from DENIS and 2MASS}
\label{section2}

The DENIS survey has produced specific observations in the direction of the bulge. \cite{Picaud2004} used these data in 94 low extinction windows with $|l|<10^\circ$ and $|b|<4^\circ$ and the Besancon population synthesis model to determine shape parameters of the bulge as well as constraints on its age and luminosity function.

\begin{figure}[ht]
\begin{center}
        \resizebox{9cm}{!}{\includegraphics  {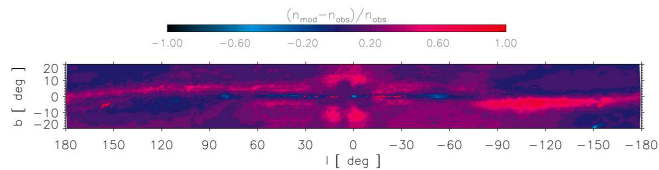}}
\caption{Comparison between predicted star counts from the Galaxy model and 2MASS data. The colours code the relative difference $(n_{mod}-n_{obs})/n_{obs}$, $n_{mod}$ and $n_{obs}$ being star counts per square degree to magnitude $K_S$=12 for the model and the data respectively. It varies from a factor of two excess in modeled star counts (in red) to a factor of two deficiency in modeled
star counts (in dark blue). 
Regions where the Galaxy model overestimates or underestimates de density of stars clearly appear.}
\label{fig1}
\end{center}
\end{figure}

Using the 3D extinction map described above and the Galaxy model thus constrained, it is possible to perform a realistic comparison of the number of stars from the Galactic model with the number of observed stars in the 2MASS data (Fig.~\ref{fig1}). The comparison points to significant discrepancies. In particular the bulge as defined by \cite{Picaud2004} discrepants from the data at $4<|b|<10$, probably because the bulge shape was determined from a more restricted area in latitudes 
A better adjustement of all 11 parameters characterizing the bulge over the whole 2MASS data set towards the bulge is ongoing.

\section{The bulge with MegaCam}
\label{section3}

We started a photometric and astrometric survey of the bulge with MegaCam at CFHT. We observed 22 fields covering $-5^\circ< l <+5^\circ$, $-1^\circ<b<+1^\circ$, in the $r'$, $i'$, and $z'$ bands. The observations are made at two epochs, separated by 3 years and have just been completed.

This survey provides unprecedented observations of the red clump very close to the Galactic plane. It will bring constraints on the luminosity function of the bulge. Moreover, a detailed extinction map with a very high spatial resolution can be obtained, using theoretical isochrones, such as the map obtained with Spitzer data (Schultheis et al., submitted). Proper motions will also be measured. We expect to get a precision of 1 mas yr$^{-1}$, corresponding to a velocity of 18 km s$^{-1}$ at the bulge distance. 6 of the fields have also Giraffe observations (PI Babusiaux). In these fields we will get the 3 components of the velocity, as well as metallicity, allowing us to investigate the kinematics/metallicity relation.

\begin{figure}[ht]
\begin{center}
        \resizebox{5.5cm}{!}{\includegraphics  {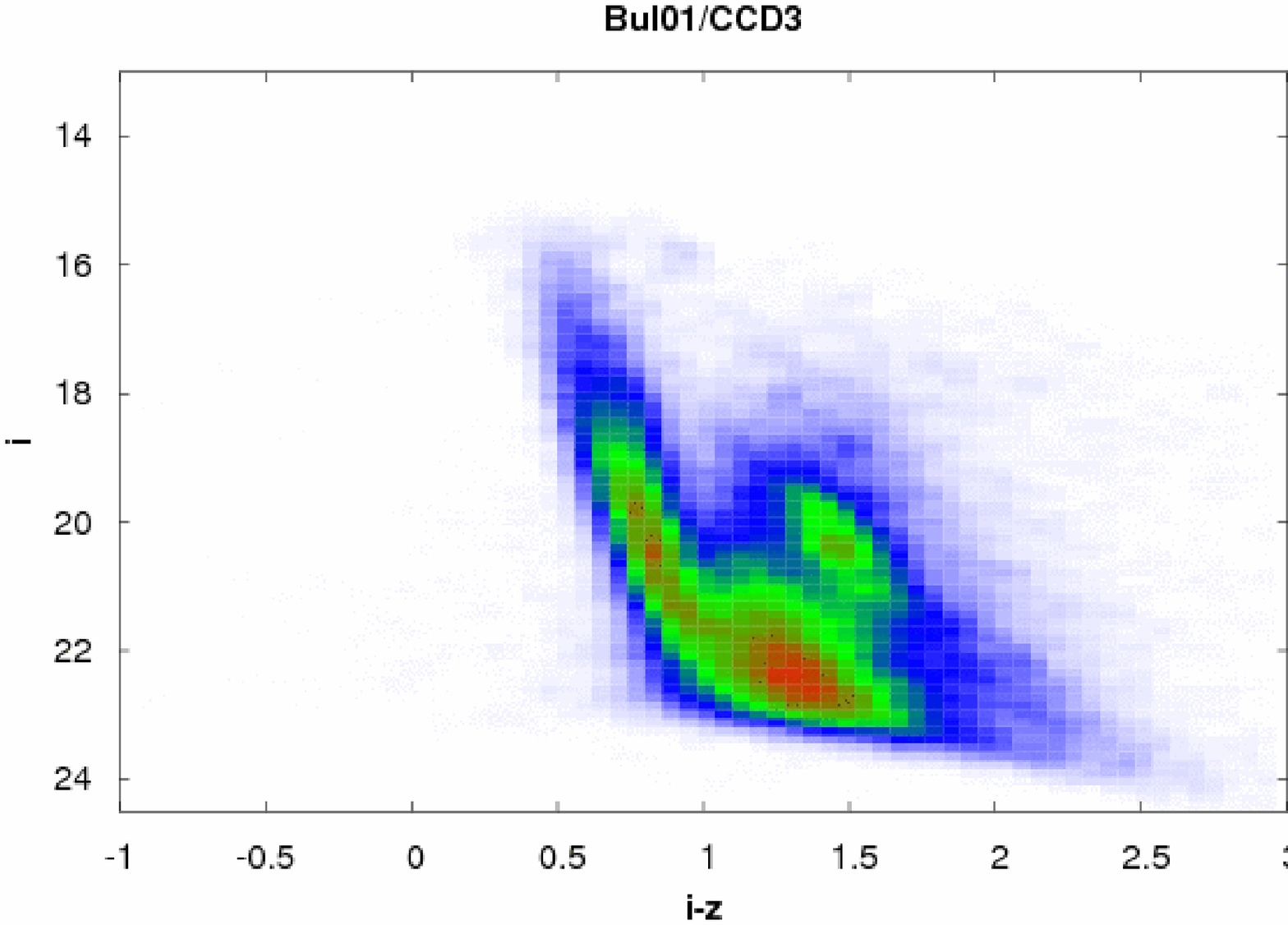}}
        \resizebox{5.5cm}{!}{\includegraphics  {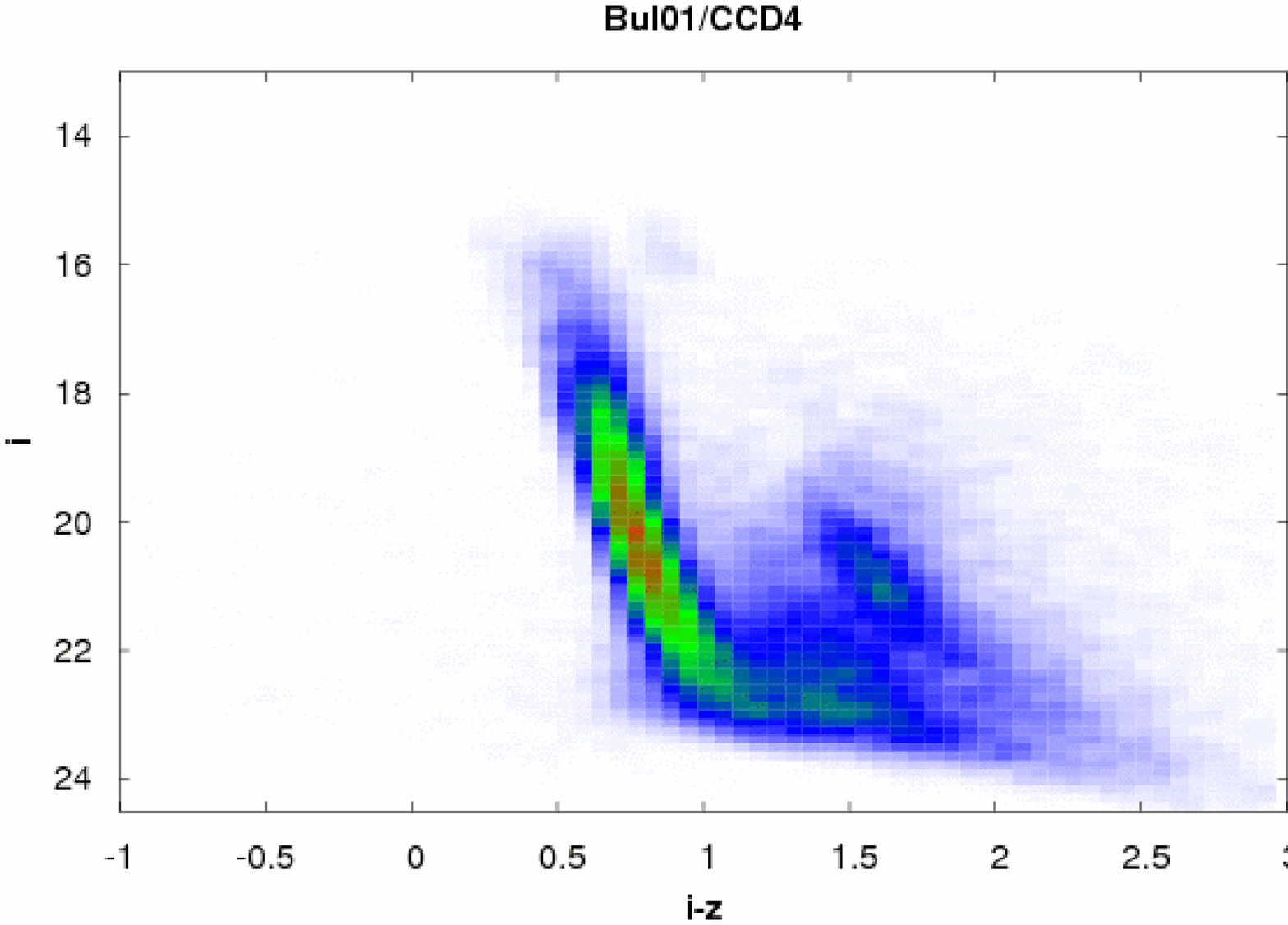}}
        \resizebox{5.5cm}{!}{\includegraphics  {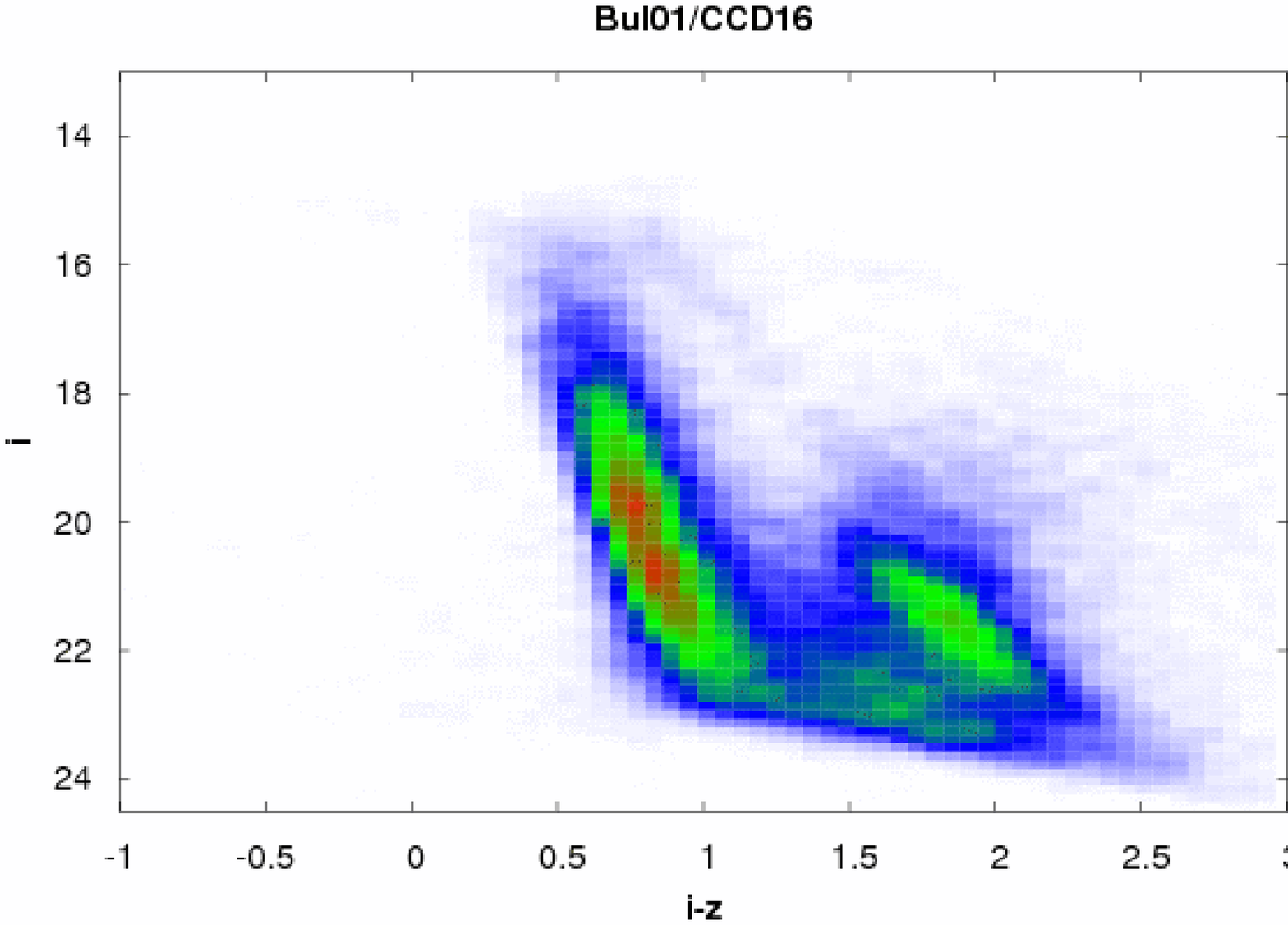}}
\caption{$i'$ versus $i'-z'$ diagram in 3 CCDs of one of the bulge fields.}
\label{fig2}
\end{center}
\end{figure}

Fig.~\ref{fig2} shows the colour-magnitude diagram for one field, in different parts of the CCDs mosaic. It shows how patchy and variable the extinction is. The red clump is clearly visible in some parts of the image, whereas it is not detected in other more extinguished parts.

\section{The bulge with Gaia}
\label{section4}
 
Gaia  will provide accurate positional, radial velocity, and photometry measurements for 1 billion of stars, that is about 1\% of the Milky Way stellar content.  Each star will be observed about one hundred times during the time of the mission, allowing the determination of proper motions.
It comprises an astrometric instrument, photometers (GRP and GBP) and a spectrometer (RVS). The limit magnitude is $G$=20\footnote{$G$ is the photometric band used by the GAIA sky mappers. It is close to the $V$ magnitude for stars with $V-I = 0$}, except for the RVS ($G$=17).
Detection problems may occur for these low spatial resolution instruments if the observed field is crowded. The estimated crowding limit is 600 000 stars deg$^{-2}$ on the astrometric fields and photometers, and 40 000 stars deg$^{-2}$ on the spectrometer.

Observations of bulge stars will strongly depend on the extinction and on the crowding.
If the extinction is too high, the number of stars will be
low (no crowding) but conversely the bulge stars would be out of reach, as they would be too faint. If the extinction
is low (like in Baade's window), bulge giants on the red clump 
are bright enough
to be reached,
but the crowding will limit the number and/or the quality of their 
measurements. Of course the number of stars in the bulge also strongly 
depends on latitudes and longitudes.

\begin{figure}[ht]
\begin{center}
        \resizebox{6.5cm}{!}{\includegraphics  {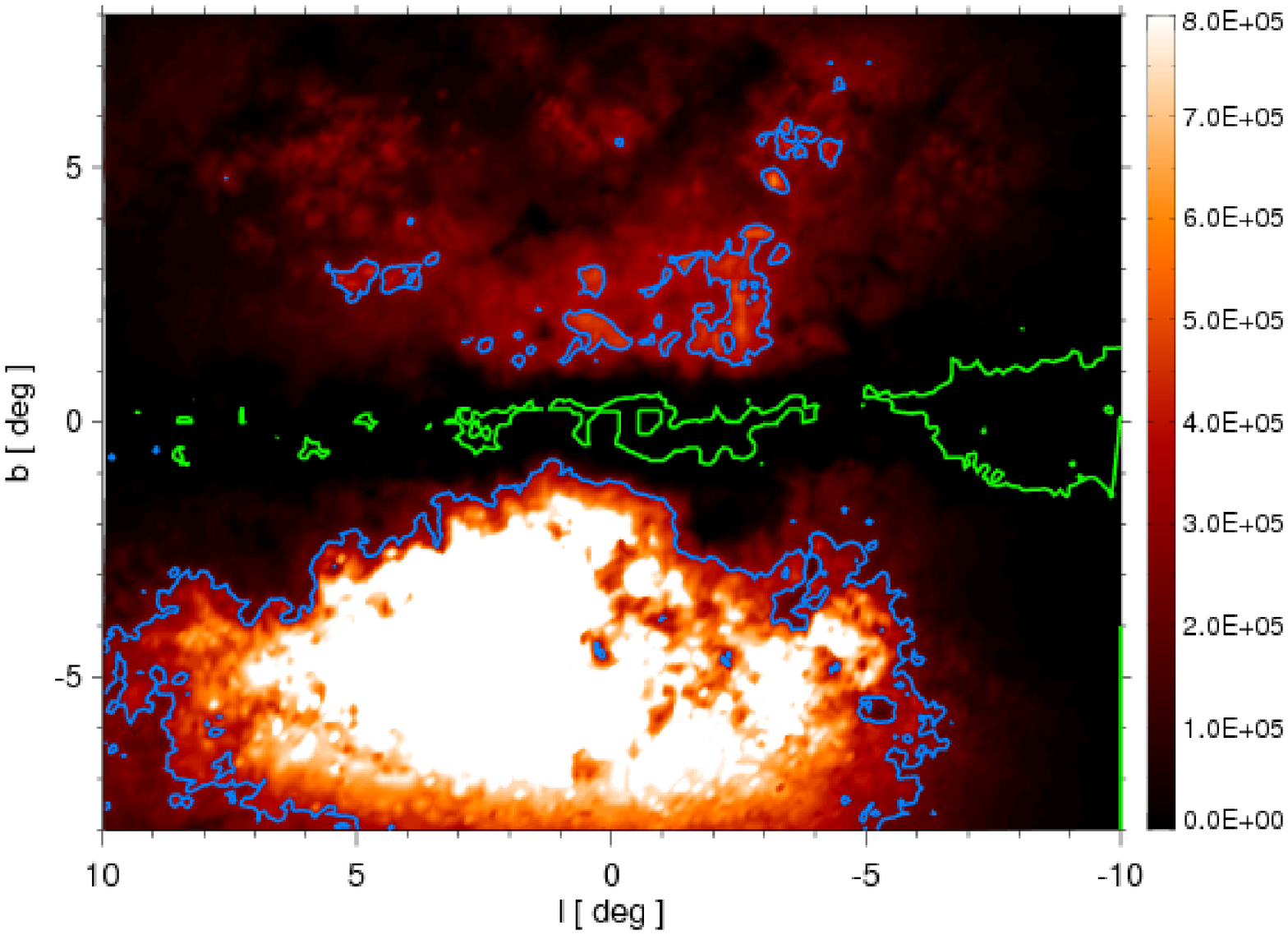}}
        \resizebox{6.5cm}{!}{\includegraphics  {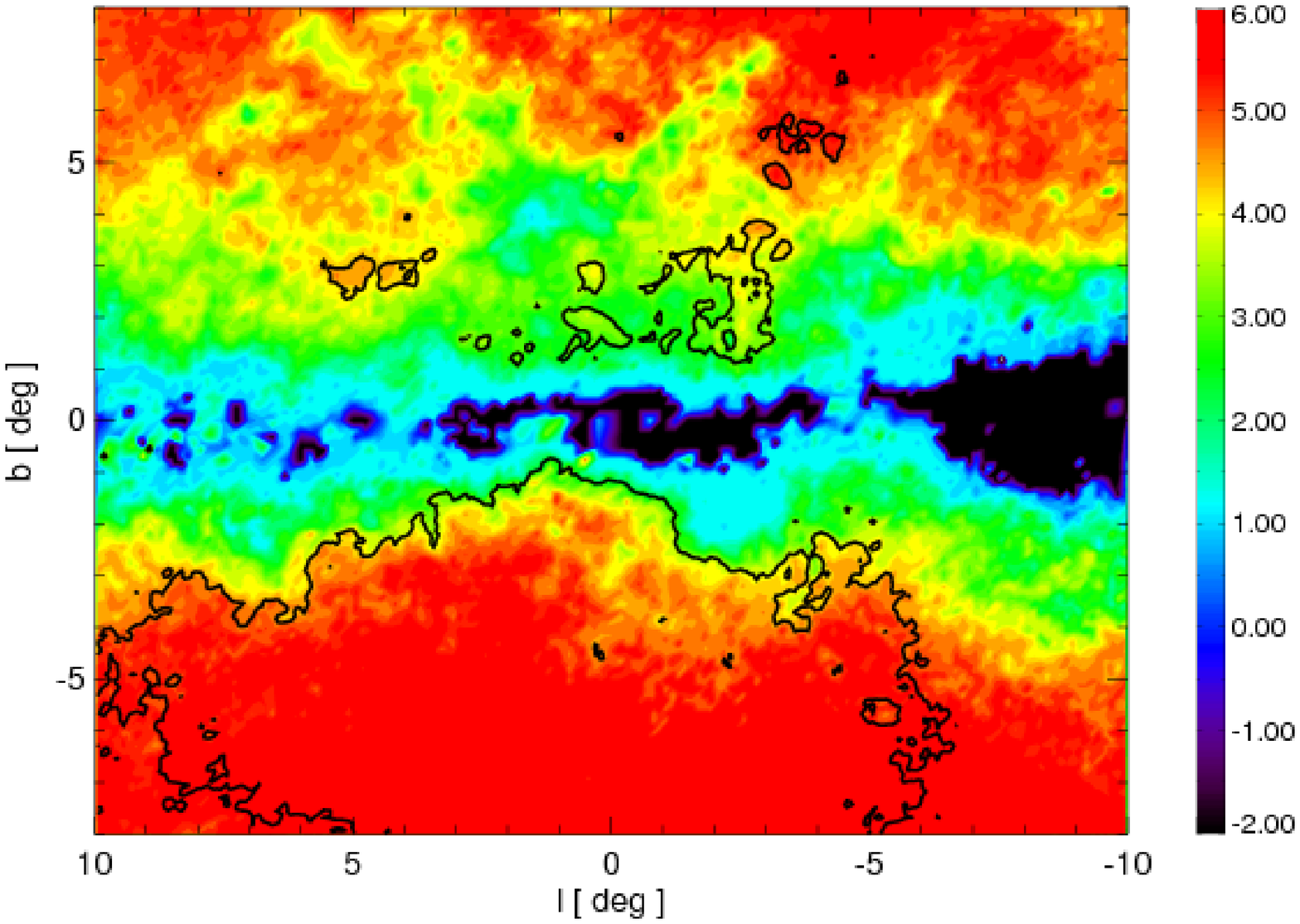}}
\caption{Left: Density of the bulge in the Gaia astrometric fields at the limiting magnitude $G$=20, as a function
of latitude and longitude. The bleu contour shows the iso-density of 600 000 stars deg$^{-2}$, which is the crowding limit in the astrometric fields. The green contour shows the iso-density of 100 stars deg$^{-2}$.
Right: Absolute magnitude M$_V$ of bulge stars just reached at the limiting magnitude $G$=20.}
\label{fig3}
\end{center}
\end{figure}

\begin{figure}[ht]
\begin{center}
        \resizebox{6.5cm}{!}{\includegraphics  {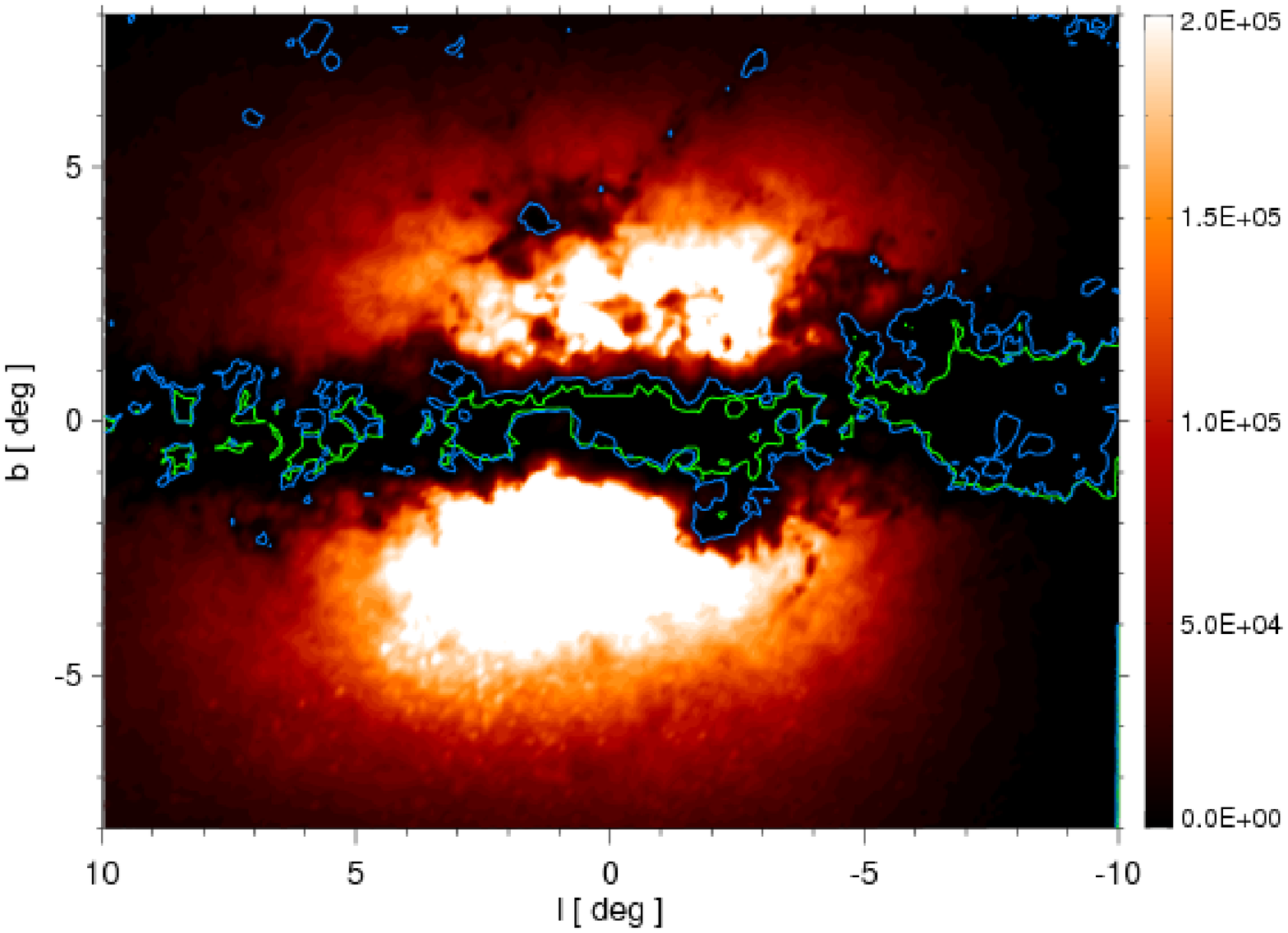}}
        \resizebox{6.5cm}{!}{\includegraphics  {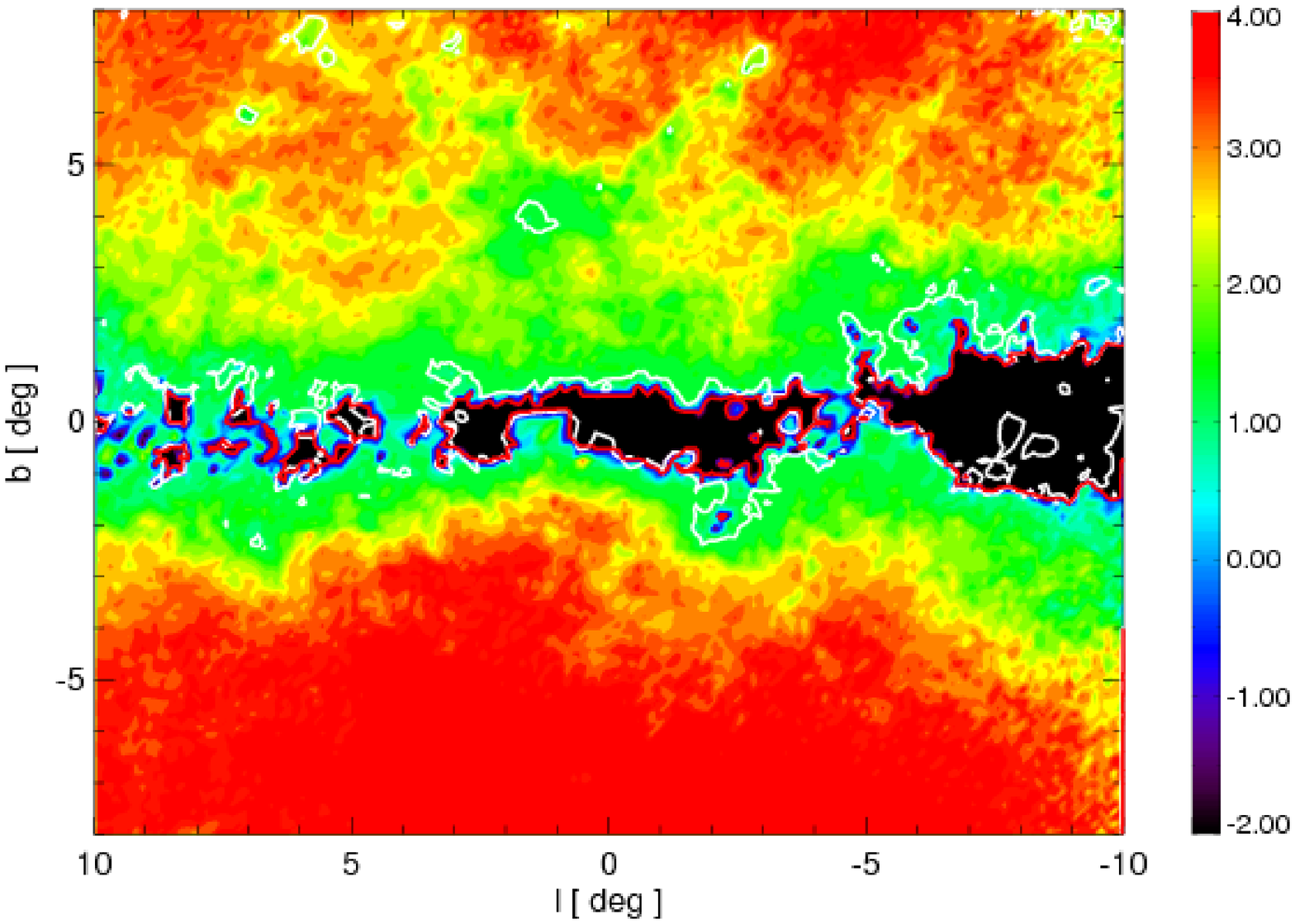}}
\caption{Same as Fig.~\ref{fig3} for the RVS (limiting magnitude $G$=17 and crowding limit of 30 000 stars deg$^{-2}$).}
\label{fig4}
\end{center}
\end{figure}

We use the Galaxy model together with the \cite{Marshall2006} extinction map to address the following question: is there a combination of 
parameters (extinction, latitude) in the galactic bulge where the extinction is
large enough to avoid crowding and not too high to allow bulge star 
measurements with Gaia instruments? 
The results of the simulations are shown in Fig.~\ref{fig3} for the astrometric fields and in Fig.~\ref{fig4} for the spectrometer. The left panel shows the density of bulge stars at the limiting $G$ magnitude of the instrument, as a function of longitude and latitude. The blue contour depicts the density at which crowding occurs. The green contour shows a iso-density of 100 bulge stars deg$^{-1}$, value at which we consider that the number of bulge stars starts to be significant. The right panel shows the absolute magnitude $M_V$ of the intrinsically faintest bulge stars reached at the limiting $G$ magnitude.

In the astrometric fields, a large part of the bulge will be visible, mainly in the Northern hemisphere, where extinction is higher. That corresponds to a number of 23 million bulge stars over an area of 220 deg$^{2}$. 
Even still, in the Southern hemisphere bulge stars brighter than the limiting magnitude $G$=20 will be detected.
Turn-off stars  ($M_V \sim 4$) and even main sequence stars are observable at high latitudes. At lower latitudes, including regions very close to the Galactic plane, the absolute magnitude of the bulge stars is $M_V \sim$ 1 to 2, and these stars are mainly clump giants. 
In the spectrometer (Fig.~\ref{fig4}) where the crowding is a much more dramatic issue, there are unextended regions where the extinction is high enough to make the crowding low but still not too strong to mask competely bulge stars. About 30 000 bulge stars over 9.7 deg$^2$ are predicted to be observed, in regions around $b =$ 1 to 2$^\circ$. They are clump giants.

There will be opportunities for Gaia to access reliable measurements
in the Galactic bulge, photometry and astrometry, parallaxes, and proper motions, in nearly all the bulge regions.These regions  strongly
depend on the assumed extinction. However, if the extinction maps do not suffer
from systematics, there should be accessible windows well spread in longitude quite
close to the dust lane in the Galactic plane. Observable bulge stars will also be spread in depth well inside
the bulge.
At low latitudes in the dense dust lane, most of the bulge stars will be too faint
(due to extinction) to be reached. However in a few fields close to
the Galactic plane in windows of lower extinction, stars in the giant clump should be
observed. In any event, the position of the fields given here
should be taken with caution because of the extreme sensitivity of the computation
to the extinction which is not known to better than about 2 magnitudes in the most
obscured regions.
Putting together observations of the different instruments including RVS, Gaia
should produce a detailed survey of bulge giants in terms of photometry as well
as kinematics. Detailed analysis of these data sets should allow us to put strong
constraints on the bulge structure and history.

\section{Conclusion}

Thanks to new photometric and astrometric surveys, the bulge structure can be revealed, bringing important constraints on the galaxy formation and evolution scenarii. We presented here several studies using such data set. Other available data are worthwhile to be analysed, such as the 49 OGLE-II fields for which proper motions have been measured \cite{Sumi}. A preliminary analysis of these data with the Galaxy model shows that the bulge rotation can be constrained with these data.

Furthermore, radial velocities and metallicities data are available in the bulge directions \citep{Ibata1995,Minniti1996,Tiede1999,Rich}.The analysis of these data will allow us to refine the kinematical parameters of the bulge, as well as bulge metallicity.
Finally, we showed that Gaia should give a detailed survey of bulge stars, down to the main sequence, in terms of photometry but also kinematics, including regions very close to the Galactic plane..


\end{document}